\documentclass[twocolumn,english,prb,epsfig,rotate,showpacs,aps,longbibliography]{revtex4-2}
\usepackage[T1]{fontenc}
\usepackage[latin9]{inputenc}
\usepackage{color}
\usepackage{verbatim}
\usepackage{amsmath}
\usepackage{amssymb}
\usepackage{graphicx}
\usepackage{esint}
\usepackage{microtype}
\usepackage{hyperref}
\usepackage{lmodern}
\makeatletter
\usepackage{babel}
\usepackage{titlesec}
\usepackage{epsfig}
\usepackage{amsmath}
\usepackage{graphicx}
\usepackage{dcolumn}
\usepackage{bm}
\usepackage{amsfonts,amssymb}
\usepackage[T1]{fontenc}
\usepackage[latin9]{inputenc}
\usepackage{amsmath}
\usepackage{graphicx}
\usepackage{amssymb}
\usepackage{esint}
\usepackage{babel}
\usepackage{microtype}
\usepackage{xcolor}
\usepackage{ulem}
\newcommand{\blue}{\color{black}}
\newcommand{\red}{\color{black}}
\newcommand{\bluee}{\color{black}}
\begin{document}
\title{Enhanced Superconducting Diode Effect in the Asymmetric Hatsugai-Kohmoto Model}
\author{Kai Chen$^{\dagger}$}
\affiliation{School of Physics Science and Engineering, Tongji University, 200092 Shanghai, China}
\author{Pavan Hosur$^{*}$}
\affiliation{Department of Physics and Texas Center for Superconductivity, University of Houston, Houston, TX 77204}
\date{\today}
\begin{abstract}
The superconducting diode effect (SDE), characterized by a nonreciprocal supercurren, has attracted significant attention in recent years due to its potential applications. However, most studies have focused on weakly correlated models, leaving the impact of strong electron-electron interactions on the SDE largely unexplored. In this work, we bridge this gap by investigating the SDE in asymmetric band metals with Hatsugai-Kohmoto (HK) interaction, which are exactly solvable due to their locality in Bloch momentum space. Through a combination of low-energy analysis and a numerical self-consistent approach, we demonstrate that HK interaction can enhance the SDE's quality factor. Our findings shed light on the role of strong electron-electron correlations in shaping the SDE.
\end{abstract}
\maketitle

\section{Introduction}
Nonreciprocity, the phenomenon where the response of a system depends on the direction of an applied stimulus, lies at the heart of modern electronic and photonic device functionality \cite{tokura2018nonreciprocal,caloz2018electromagnetic}. One prominent manifestation of nonreciprocity is the rectification effect in semiconductor diodes, where current flows preferentially in one direction, enabling crucial applications such as signal modulation and logic operations. The foundational principle behind this behavior is the breaking of inversion symmetry in the system, which generates an asymmetry in charge carrier dynamics under an external bias.

Recent advances have unveiled analogous nonreciprocal effects in superconducting systems, referred to as the SDE. The SDE is characterized by a non-reciprocal critical supercurrent, where dissipationless current flows preferentially in one direction but is suppressed in the opposite direction \citep{ando2020observation, nadeem2023superconducting, ma2025superconducting, fukaya2025superconducting}. This phenomenon extends the concept of rectification into the superconducting regime, presenting significant potential for energy-efficient superconducting electronics. In contrast to semiconductor diodes, the SDE arises from the simultaneous breaking of time-reversal and inversion symmetries \citep{nagaosa2024nonreciprocal}, achieved through mechanisms such as intrinsic spin-orbit coupling, external magnetic fields, chiral structures, nontrivial surface states, or finite-momentum Cooper pairing \citep{yuan2022supercurrent, he2022phenomenological, daido2022intrinsic, fulde1964superconductivity, larkin1965nonuniform, legg2022superconducting, ilic2022theory, yuan2023surface, banerjee2023phase, he2023supercurrent}. Notably, superconducting states with finite-momentum Cooper pairing, analogous to the Fulde-Ferrell-Larkin-Ovchinnikov state \citep{fulde1964superconductivity, larkin1965nonuniform}, are believed to play a central role in realizing the SDE.

Experimental evidence for the SDE has been observed in a variety of systems, including magic-angle twisted graphene (TBG)\citep{lin2022zero,scammell2022theory,diez2023symmetry}, van der Waals heterostructures \citep{bauriedl2022supercurrent, wu2022field}, topological insulator/superconductor hybrids \citep{yasuda2019nonreciprocal, masuko2022nonreciprocal, karabassov2022hybrid,anh2024large, nikodem2025tunable, kudriashov2025non,nagahama2025two}, the noncentrosymmetric superconductor T$_d$-MoTe$_2$ \cite{du2024superconducting}, superconductor layers with EuS \cite{hou2023ubiquitous}, superconducting thin films patterned with conformal-mapped nanoholes \cite{lyu2021superconducting}, Josephson junction arrays \citep{baumgartner2022supercurrent,pal2022josephson,gupta2023gate,ghosh2024high,chen2024edelstein,kim2024intrinsic,turini2022josephson,zhang2024magnetic,zhao2023time,trahms2023diode,he2024observation,cayao2024enhancing,mondal2025josephson,ingla2025efficient, borgongino2025biharmonic,schulz2025quantum,kudriashov2025non,ma2025field}, superconductor nanowire/topological Dirac semimetal hybrid systems \citep{ishihara2023giant, zhang2025realizing,  travaglini2025robust, ge2025nonreciprocal, mayo2025band}, magnetic-field-free SDEs in FeSe \cite{nagata2025field}, high-T$_c$ cuprates \cite{qi2025high}, and heterostructures of Ising superconducting NbSe$_2$ and ferromagnetic Fe$_3$GeTe$_2$ \cite{hu2025tunable}.

These experimental advancements have stimulated extensive theoretical efforts to understand the mechanisms underlying the SDE. Key focuses include exotic Cooper pairing \cite{yuan2022supercurrent,daido2022intrinsic,he2023supercurrent}, coupling between supercurrent and symmetry-breaking order parameters \cite{banerjee2024enhanced}, competition between pairing channels \cite{chen2024intrinsic}, competition between finite-momentum and zero-momentum superconducting states \cite{chakraborty2024perfect}, geometric properties of electronic wavefunctions \cite{schulz2025quantum,hu2025geometric}, and other theoretical frameworks for understanding the SDE \cite{hu2007proposed,zhang2022general,davydova2022universal,mao2024universal,virtanen2024nonreciprocal}. However, most theoretical studies have primarily focused on weakly correlated electron systems, largely overlooking the role of strongly correlated effects \cite{zhuang2025current,sim2025pair,bhowmik2025optimizing,yuan2025orbital,hasan2025superconducting,costa2025unconventional}.
 
Although SDEs have been observed in strongly correlated systems, such as moir\'e materials \cite{lin2022zero,diez2023symmetry}, the role of electron correlations in the SDE remains an open question. These correlations are known to be central to unconventional superconductivity \cite{cao2018unconventional,kerelsky2019maximized,liu2021tuning} and other quantum phenomena \cite{nuckolls2020strongly,choi2019electronic,kouwenhoven2001revival}. However, understanding their impact on the SDE is challenging, largely due to the complexity of handling strong electron-electron interactions.

The HK model \cite{hatsugai1992exactly,baskaran1991exactly,hatsugai1996mutual,zhao2025hatsugai}, with its momentum-localized interaction and exact analytical solvability, provides a compelling framework for exploring the effects of strong electron-electron correlations on the SDE. The HK model is a cornerstone for studying non-Fermi liquid behavior, characterized by non-Landau quasiparticle excitations that violate Luttinger's theorem \cite{baskaran1991exactly,phillips2020exact,zhong2022solvable}. These excitations result in a Green's function resembling the Yang-Rice-Zhang ansatz for cuprates \cite{yang2006phenomenological,rice2011phenomenological}, where zeros form a Luttinger surface instead of a conventional Fermi surface \cite{dzyaloshinskii2003some,dave2013absence}. This property underscores the manifestation of Mottness in the strong-coupling regime and unconventional metallic behavior at weaker couplings \cite{huang2022discrete}. Moreover, investigations of the HK model have revealed Cooper pairing instabilities and dynamic spectral weight transfer \cite{setty2020pairing}. {\bluee{In the context of the symmetric band structure, the interplay of the HK interaction and BCS pairing has been shown to fundamentally alter the nature of the superconducting state. Unlike the continuous second-order transition in standard BCS theory, strong momentum-space correlations in the HK model drive the superconducting phase transition to become first-order, separated from the weakly-correlated regime by a tricritical point. Furthermore, the symmetric model exhibits anomalous thermodynamic properties, including an enhanced condensation energy, the absence of the Hebel-Slichter peak in NMR relaxation \cite{zhao2022thermodynamics}, and a remarkable two-stage superconductivity, where the order parameter undergoes a second sudden jump at a lower temperature due to the release of residual entropy from the singly-occupied momentum region \cite{li2022two}.}}

In this work, we explore the SDE in an asymmetric band metal (ABM), a metal with a skewed dispersion satisfying $E\left(k\right) \neq E\left(-k\right)$, coupled with HK interaction. ABMs break time-reversal and inversion symmetries, and in the weakly correlated regime they give rise to striking behaviors, such as a perfect superconducting diode effect when proximate to a conventional superconductor \citep{hosur2022equilibrium}, {\blue{nonoreciprocal superconductivity with characteristic signatures in Andreev reflection \cite{davydova2024nonreciprocal},}} and a new regime of localization under disorder \citep{arora2023suppression}. The introduction of HK interaction is expected to alter the electronic band structure, Fermi surface geometry, and pairing momenta, which could influence the SDE, but the precise interplay between these factors remains unclear. By combining low-energy effective theory with self-consistent numerical simulations, we demonstrate that strong electron correlations provide a powerful mechanism for amplifying the nonreciprocal supercurrent. {\red{This enhancement originates from a correlation-induced band splitting, which causes the superconducting spectra associated with the resulting subbands to become gapless at distinct Cooper pair momenta.}} Our findings offer fundamental insights and a potential design principle for optimizing the superconducting diode effect in correlated materials.

This paper is organized as follows. We begin in Section \ref{introHK} by elucidating the influence of the HK interaction on the ABM band structure. Section \ref{HK-low-energy} then develops a low-energy theory to account for the SDE. The dependence of the SDE quality factor on the HK interaction strength is numerically established in Section \ref{HK-model}. We present our concluding remarks in Section \ref{CONHK}.

\section{HK model with asymmetric band dispersion}
\label{introHK}
In the normal phase, the HK model exhibits non-Fermi liquid behavior \cite{phillips2020exact,zhao2023friedel} and in the superconducting phase with sufficiently large HK interaction, this model undergoes a first-order phase transition as the temperature crosses the critical temperature \cite{zhao2022thermodynamics,li2022two}. On the other hand, the so-called ABM state in the superconducting phase exhibits SDEs. How does the HK interaction affect the SDEs in the superconducting ABM? To answer this question, let's consider the following model:
\begin{equation}
H\left(p\right)=\sum_{p,\sigma=\uparrow\downarrow} \xi_p \hat{n}_{p,\sigma} + U \sum_{p} \hat{n}_{p,\downarrow}\hat{n}_{p,\uparrow},
\label{H_HK0}
\end{equation}
where $\hat{n}_{p,\sigma} \equiv \hat{c}_{p\sigma}^{\dagger} \hat{c}_{p\sigma}$ represents the number operator for Bloch momentum $p$ and spin $\sigma$, with $\hat{c}_{p\sigma}$ and $\hat{c}_{p\sigma}^{\dagger}$ denoting the annihilation and creation operators, respectively. {\color{black} The band dispersion is given by $\xi_p = -2 t_1 \cos(p) + t_2 \sin \left( 2p + \frac{\pi}{3} \right) - \mu$, where $\mu$ is the chemical potential, while $t_1=2 t_2$ and $t_2=1$ (which sets the energy scale in our model) are the nearest-neighbor and next-nearest-neighbor hopping amplitudes, respectively. Physically, the asymmetric character of the band ($\xi_p \neq \xi_{-p}$) arises specifically from the $\frac{\pi}{3}$ phase factor in the next-nearest-neighbor hopping term, which explicitly breaks spatial inversion symmetry.} Since $\xi_p \neq \xi_{-p}$, the dispersion is an asymmetric function of $p$, giving rise to the name asymmetric HK model. The last term in Eq. (\ref{H_HK0}) represents the HK interaction, with a strength $U > 0$, corresponding to an on-site repulsion in momentum space. Because the HK interaction is local in momentum space, this model is exactly solvable, and the single-particle Green's function is given by \cite{nogueira1996study,phillips2020exact,zhao2022thermodynamics,li2022two,bacsi2025infinite}:
\begin{figure}[h]
    \centering
    \includegraphics[width=1\linewidth]{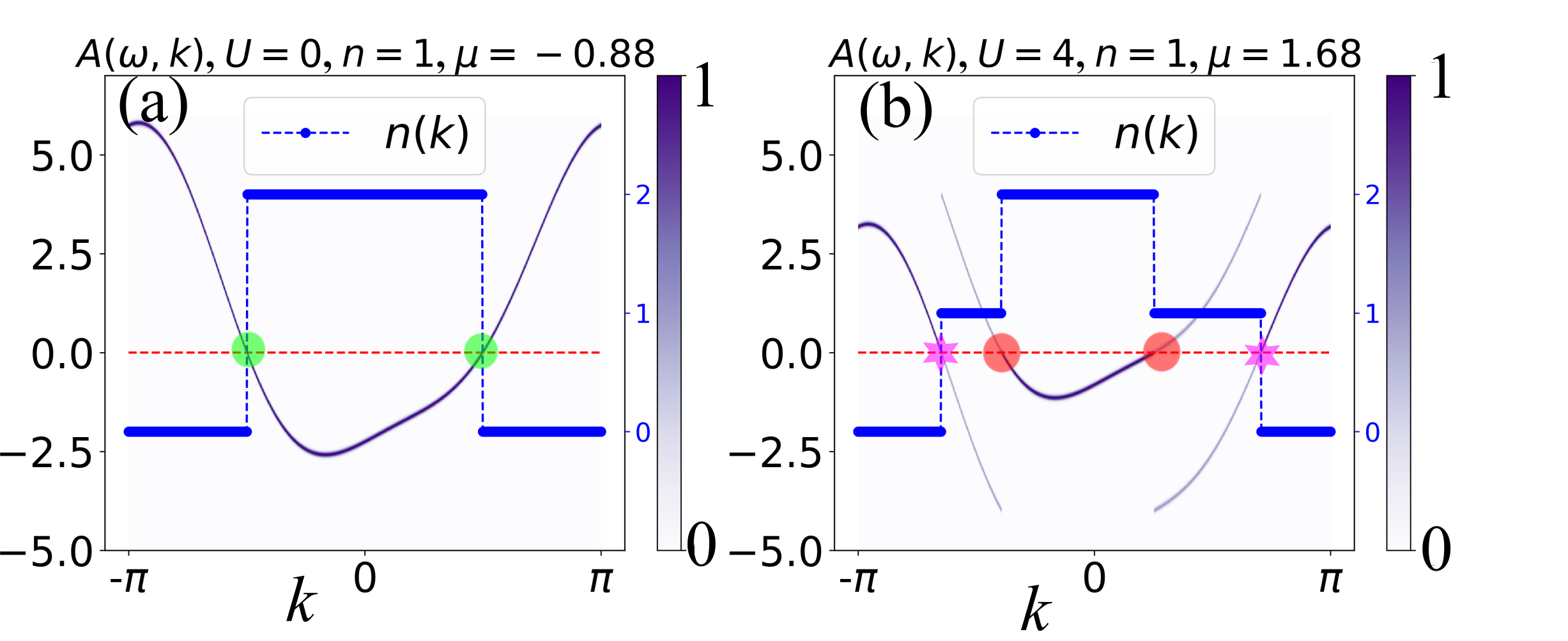}
   \caption{Properties of the asymmetric HK model. (a) Spectral function $A(\omega,k)$ and particle number distribution $n(k)$ at $U=0$. The Fermi momenta $k_{F\pm}$ are indicated by green points. (b) Spectral function and particle number distribution at $U=4$. The Fermi momenta $k_{F\pm}^{(1)}$ and $k_{F\pm}^{(2)}$ are marked by pink stars and red points, respectively.}
    \label{hkband0}
\end{figure}
\begin{equation}
G_{\sigma}\left(i\omega_n, p\right)=\frac{1-\langle\hat{n}_{p\Bar{\sigma}}\rangle}{i\omega-\xi_p}+\frac{\langle\hat{n}_{p\Bar{\sigma}}\rangle}{i\omega-\xi_p-U},
\label{Green}
\end{equation}
with $\langle \hat{n}_{p\bar{\sigma}} \rangle$ denoting the average number for spin $\bar{\sigma} = -\sigma$ in the following ground states:
\begin{equation}
|G\rangle=\prod_{p\in\Gamma_2} \hat{c}_{p\uparrow}^{\dagger} \hat{c}_{p\downarrow}^{\dagger}\prod_{q\in\Gamma_1}\frac{1}{\sqrt{2}}\left(\hat{c}_{q\uparrow}^{\dagger} +\hat{c}_{q\downarrow}^{\dagger}\right)|\emptyset\rangle,
\label{GS}
\end{equation}
where $|\emptyset\rangle$ is the empty state. Excitations in the lower band are created by the operators $\zeta_{k\sigma}^{\dagger}=c_{k\sigma}^{\dagger}\left(1-n_{k\bar{\sigma}}\right)$. However, the two-particle excitation generated by $\zeta_{k\sigma}^{\dagger}\zeta_{k\bar{\sigma}}^{\dagger}$ equals zero. This means that some excitation states in the HK model cannot be accessed through single-particle operators. As a result, the physics of the HK model differs from that of a Fermi liquid \cite{phillips2020exact,zhao2022thermodynamics}.

At zero temperature, the particle number distribution as a function of momentum $p$ and spin $\sigma$ is given by: 
\begin{equation}
n_{p,\sigma} \equiv \langle\hat{n}_{p,\sigma}\rangle= \frac{1}{2}\left(\Theta\left(-\xi_p\right)+\Theta\left(-\xi_p-U\right)\right).
\label{number}
\end{equation}
which is spin-independent. Here, $\Theta(x)$ is the Heaviside step function.

{\bluee{Before introducing the superconducting pairing ($g=0$), it is instructive to clarify the normal state phase diagram at half-filling ($n=1$). The HK interaction exactly splits the bare band into a lower and an upper Hubbard band. A phase transition from a correlated metal to a Mott insulator occurs only when $U$ exceeds the bare bandwidth $W$, at which point a true charge gap opens between the Hubbard bands. For our asymmetric dispersion, the bare bandwidth is $W = \max(\xi_p) - \min(\xi_p) \approx 8.4$. Because the maximum interaction strength considered in our subsequent calculations is $U=6$ (which is strictly less than $W$), the normal state of our system remains a correlated metal prior to the onset of superconductivity.}}

In this work, the filling is fixed at $n = 1$. The chemical potential depends on the filling $n$ and the repulsive strength $U$, and is determined by the condition $n = \int_{k}\left(n_{k,\uparrow}+n_{k,\downarrow}\right)\equiv  \int_{k} n_{k}$.
The spectral function $A(\omega,k)$ and the particle number distribution $n_k$ are shown in the Fig. (\ref{hkband0}) for different strengths of the repulsive HK interaction. 

For a system with asymmetric band structures, entering the superconducting state can cause Cooper pairs to acquire a finite center-of-mass momentum $q$ in equilibrium. In such a state, the supercurrent $J_q$ vanishes when $q \approx k_{F+} + k_{F-}$, where $k_{F\pm}$ denote the Fermi momenta. As shown in Fig.~\ref{hkband0}, the HK interaction splits the spin-degenerate asymmetric bands into two Hubbard bands separated by an energy $U$. This interaction also leads to a splitting of the Fermi momenta $k_{F\pm}$ into four distinct values, labeled $k_{F\pm}^{(1)}$ and $k_{F\pm}^{(2)}$ (see Fig.~\ref{hkband0}), due to Fermi surface reconstruction. As a result, there exist two Cooper pair momenta, $q^{(1)} \approx k_{F+}^{(1)} + k_{F-}^{(1)}$ and $q^{(2)} \approx k_{F+}^{(2)} + k_{F-}^{(2)}$, at which the supercurrent vanishes: $J_{q^{(1)}} = J_{q^{(2)}} = 0$. One therefore expects the supercurrent to exhibit a more complex profile, which will influence the SDE. To systematically explore the effect of this band splitting on the SDE, we now turn to a low-energy description.
 
\section{Low-energy theory} 
\label{HK-low-energy}

A metal with an asymmetric band structure, when proximitized by a superconductor, can exhibit a finite and potentially perfect SDE \citep{hosur2022equilibrium}. The flow of a supercurrent confers a finite momentum $q$ to the Cooper pairs, leading to a suppression of the superconducting gap. Since the supercurrent density scales with $q$, the critical current is ultimately set by the condition for the superconducting energy spectrum (the eigenvalues of the BdG Hamiltonian) to become gapless.
 \begin{figure}[h]
    \centering
    \includegraphics[width=1\linewidth]{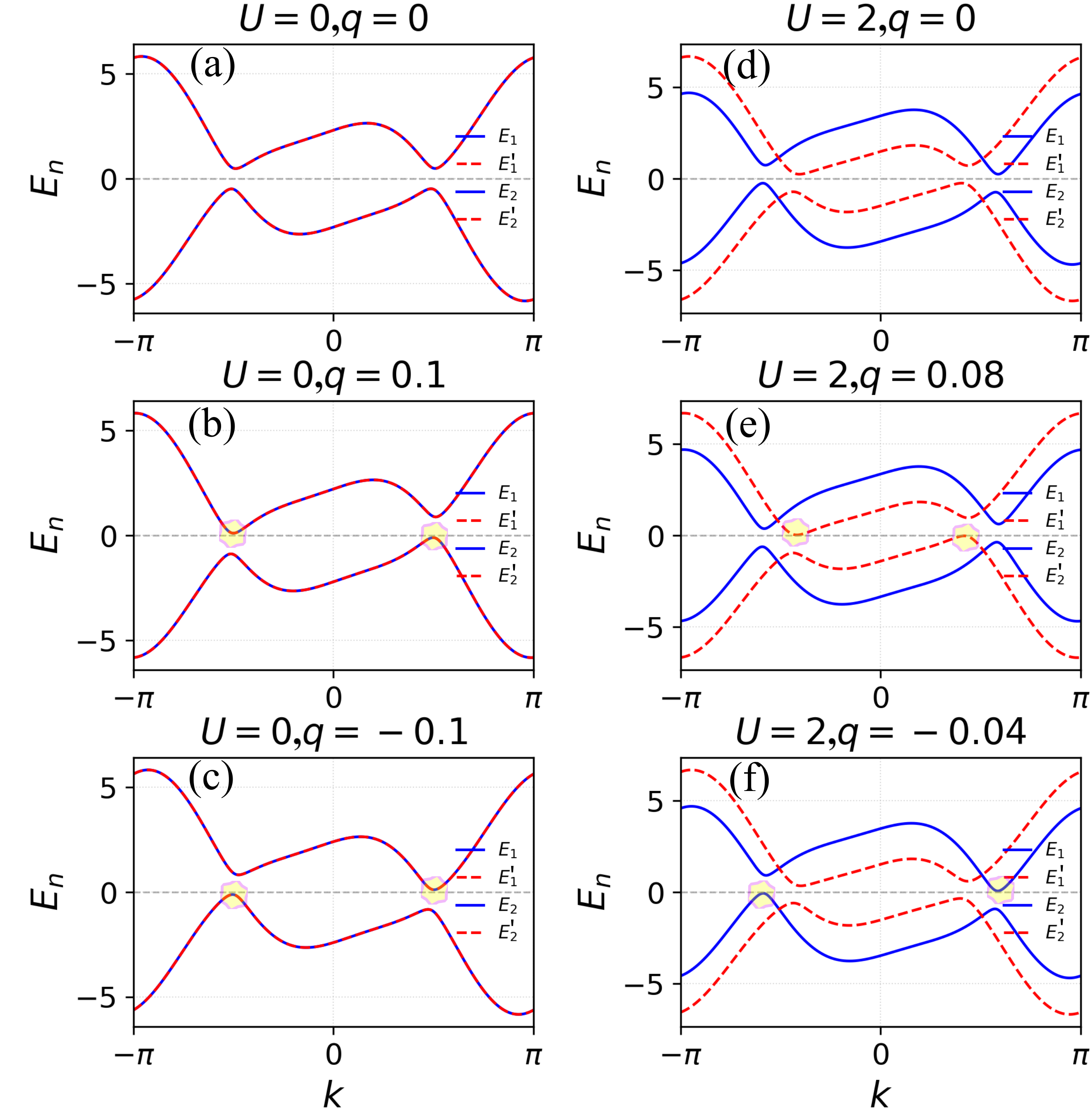}
    \caption{Superconducting energy spectra $E(k)$ for different Cooper pair momenta $q$. (a--c) $U=0$ and (d--f) $U=2$. The spectra $E_1$, $E_2$ and $E'_1$, $E'_2$ are obtained from the Bogoliubov-de Gennes (BdG) Hamiltonian with normal-state dispersions $\xi_k$ and $\xi_k+U$, respectively. Gapless points are highlighted by yellow hexagons. The pairing strength is $\Delta=0.5$.}
    \label{delta}
\end{figure}
\titlespacing*{\section}{0pt}{1cm}{0.5cm}

{\red{In our system, as shown in Fig.~\ref{hkband0}(b), the HK interaction splits the band into two subbands with energies $\xi_k$ and $\xi_k + U$, respectively. At low temperatures, only electronic states near the Fermi level are relevant for forming Cooper pairs. This corresponds to the singlet pairing between electrons indicated by the red points (for the $\xi_k+U$ subband) and the pink stars (for the $\xi_k$ subband) in Fig.~\ref{hkband0}(b).
When the Cooper pairs carry a finite center-of-mass momentum $q$, the combined effects of this finite-$q$ pairing and the HK-induced band reconstruction produce two distinct superconducting energy spectra within the BdG formalism: {\bluee{$E_n(k,q)$, which has a superconducting gap near Fermi momenta $k_{F\pm}^{(1)}$, and $E'_n(k,q)$, with a superconducting gap near Fermi momenta $k_{F\pm}^{(2)}$.}} A key consequence is that these spectra become gapless at inequivalent critical momenta, $q_{c}^+$ and $q_{c}^-$, for the two subbands. Since the supercurrent is proportional to $q$ for small momenta, this asymmetry directly results in different critical supercurrents for the two current directions, thereby generating the SDE.}}

As shown in Fig.~\ref{delta}, for $U=0$ the energy spectra close symmetrically at $q_c^\pm= \pm 0.1$. In stark contrast, for $U \neq 0$, the spectra close at two inequivalent momenta: {\bluee{$q_{c}^{-} = -0.04$ for the $E_n$ branch and $q_{c}^{+} = 0.08$ for the $E'_n$ branch}}, resulting in a pronounced asymmetry $q_{c}^{-} \neq |q_{c}^{+}|$. Given that the supercurrent is approximately proportional to $q$, this asymmetry in the spectral closing points directly implies a mismatch between the critical currents for opposing directions, thereby realizing nonreciprocal supercurrent transport---the hallmark of the SDE.

To quantify this asymmetry, {\blue{we build on the property that the supercurrent is often proportional to q, at least for small q around the equilibrum value.}} Thus, we define a parameter $\delta = \left| ( q_c^{+} - |q_c^{-}| ) / ( q_c^{+} + |q_c^{-}| ) \right|$ to characterize the disparity in critical currents. As shown in Fig.~\ref{eta}(a), $\delta$ captures the qualitative trend of the diode efficiency $\eta$, a detailed calculation of which is presented in the following section.

\titlespacing*{\section}{0pt}{1cm}{0.5cm}
\section{Numerical Calculation of the SDE in the Asymmetric HK Model}
\label{HK-model}

In the previous section, we uncovered the effects of the HK interaction on the SDE. Through the band splitting induced by the HK interaction, we propose mismatched Cooper pair momentum as a qualitative signature of the occurrence of the SDE. To uncover the interband correlation effects and elucidate how the HK interaction enhances the SDE, we now perform a fully numerical calculation of the supercurrent. We begin by introducing an $s$-wave superconducting pairing. The mean-field BdG Hamiltonian is then given by:
 \begin{eqnarray}
H_{BdG}&=&\sum_{k,\sigma=\uparrow\downarrow} \xi_k \hat{n}_{k,\sigma} + U \sum_{k} \hat{n}_{k,\downarrow}\hat{n}_{k,\uparrow} \nonumber\\
&-& \sum_{k,q}\left[\Delta_{q}\hat{c}^{\dagger}_{k+q,\uparrow}\hat{c}^{\dagger}_{-k,\downarrow} +h.c.\right] +\frac{\Delta_{q}^2}{g},
\label{H_HK_bdg}
\end{eqnarray}
where $h.c.$ denotes the Hermitian adjoint of the foregoing term in the bracket, $q$ is the momentum of Cooper pairs, and the pairing potential $\Delta_{q} \equiv g \sum_{k} \langle\hat{c}_{k+q,\uparrow} \hat{c}_{-k,\downarrow}\rangle$, with $g=1.8>0$ denoting the attractive pairing strength.

{\bluee{While standard real-space repulsive interactions (such as the Hubbard model) heavily penalize $s$-wave pairing and promote finite-angular-momentum states like $d$-wave, the HK interaction is strictly local in momentum space. Consequently, it penalizes double occupancy of specific momentum states rather than real-space sites, meaning $s$-wave pairing is not intrinsically suppressed by the geometric penalty of on-site repulsion. Furthermore, previous studies on the symmetric HK model have demonstrated that the strongly correlated phenomena induced by the HK interaction manifest almost identically in both $s$-wave and $d$-wave channels \cite{zhao2022thermodynamics}. Therefore, without loss of generality, we adopt the $s$-wave channel as a representative framework to cleanly isolate the effect of the HK interaction on the SDE.}}

In the composite Hilbert space $\mathcal{H}_{k+q} \otimes \mathcal{H}_{-k}$, which has a 16-dimensional basis $\{|\emptyset\rangle, |\uparrow\rangle, |\downarrow\rangle, |\uparrow\downarrow\rangle\}_{k+q} \otimes \{|\emptyset\rangle, |\uparrow\rangle, |\downarrow\rangle, |\uparrow\downarrow\rangle\}_{-k}$, the BdG Hamiltonian $H_{\text{BdG}}$ {\red{decomposes into 9 distinct sectors.}} {\blue{These sectors are characterized by different spin configurations of electrons with crystal momenta $k+q$ and $-k$}}. The full Hamiltonian is presented in \ref{appendix}.

\begin{figure}[h]
    \centering
    \includegraphics[width=1\linewidth]{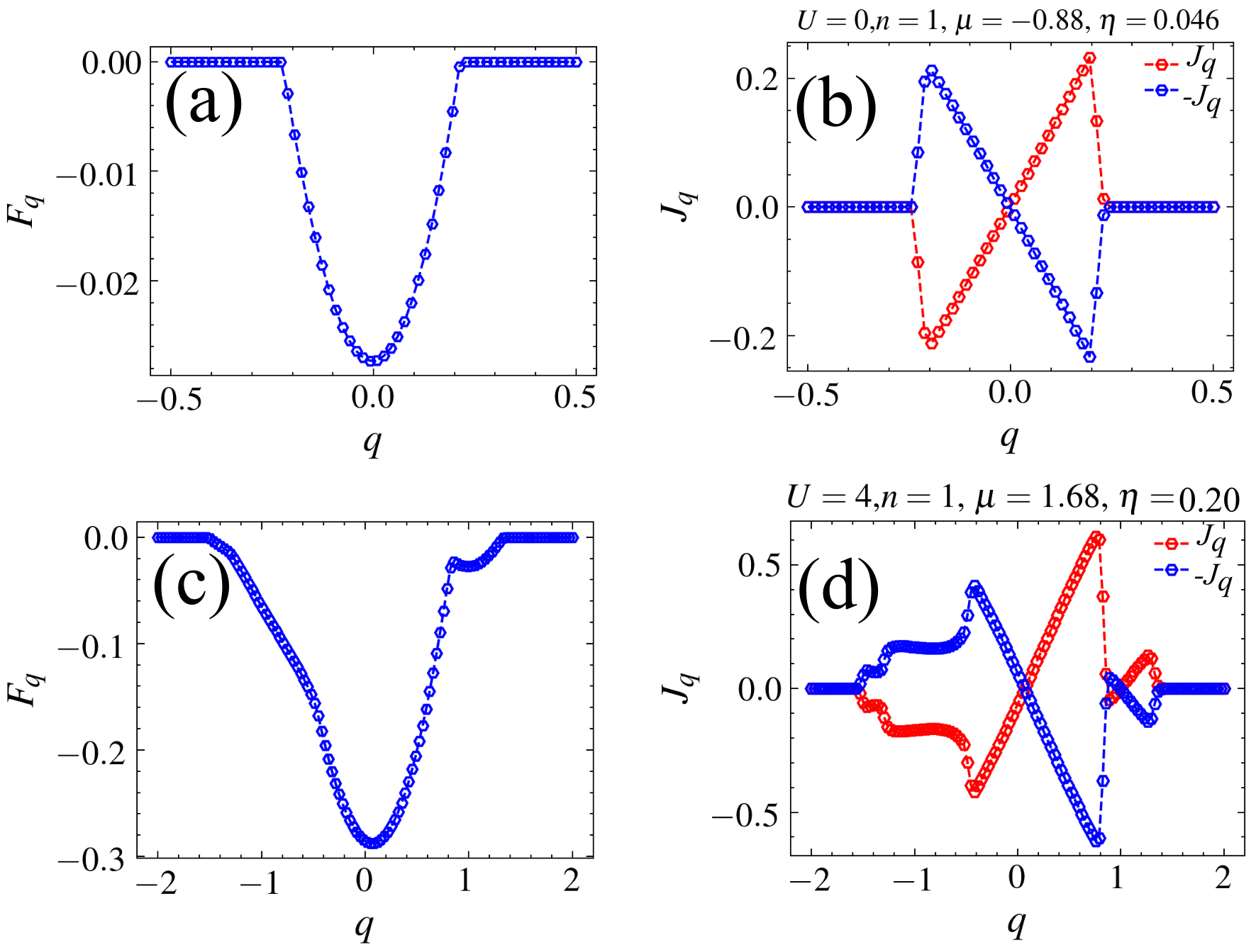}
    \caption{Condensation energy $F_q$ and supercurrent $J_q$ as functions of the Cooper pair momentum $q$. (a, b) Results for the HK interaction strength $U = 0$. (c, d) Results for $U = 4$. The temperature $T=0.03$. For all panels, the attractive pairing strength is fixed at $g=1.8$.}
    \label{freej}
\end{figure}

To obtain the superconducting current as a function of $q$, we first calculate the condensation energy at temperature $T$, which is expressed as: 
\begin{equation}
F_{q,\Delta_{q}} = -k_B T \int\frac{d^Dk}{\left(2\pi\right)^D} \ln Z_{k,q} - (\Delta_{q}=0\quad \text{contribution}).
\label{free}
\end{equation}
Here, $D$ denotes the spatial dimension, $k_B$ is the Boltzmann constant (set to 1 throughout this work), and the partition function is given by: 
\begin{equation}
Z_{k,q} = \sum_{n} e^{-\beta (E_{n,k,q}+\Delta_{q}^2/g)}, 
\label{part}
\end{equation}
with $\beta$ being the inverse temperature. The $E_{n,k,q}$ in the partition function $Z_{k,q}$ are the eigenvalues of the Hamiltonian $H_{\text{BdG}}$. Although analytical expressions for the eigenvalues can be derived, the condensation energy remains highly complex, even in this solvable model. Consequently, we employ numerical methods to minimize the condensed energy, $F_{q,\Delta_{q}} - F_{q,\Delta_{q}=0}$, to determine the optimal pairing potential $\Delta_q$. With $\Delta_q$, we compute the condensate energy and evaluate the supercurrent $J_q = 2\frac{\partial F_q}{\partial q}$ as a function of $q$. This analysis facilitates the extraction of the SDE quality factor, $\eta = \left|\frac{J_c^{+} - |J_c^{-}|}{J_c^{+} + |J_c^{-}|}\right|$, where $J_c^{\pm}$ represent the critical supercurrents in opposite directions along the momentum axis.
\begin{figure}[h]
    \centering
    \includegraphics[width=1\linewidth]{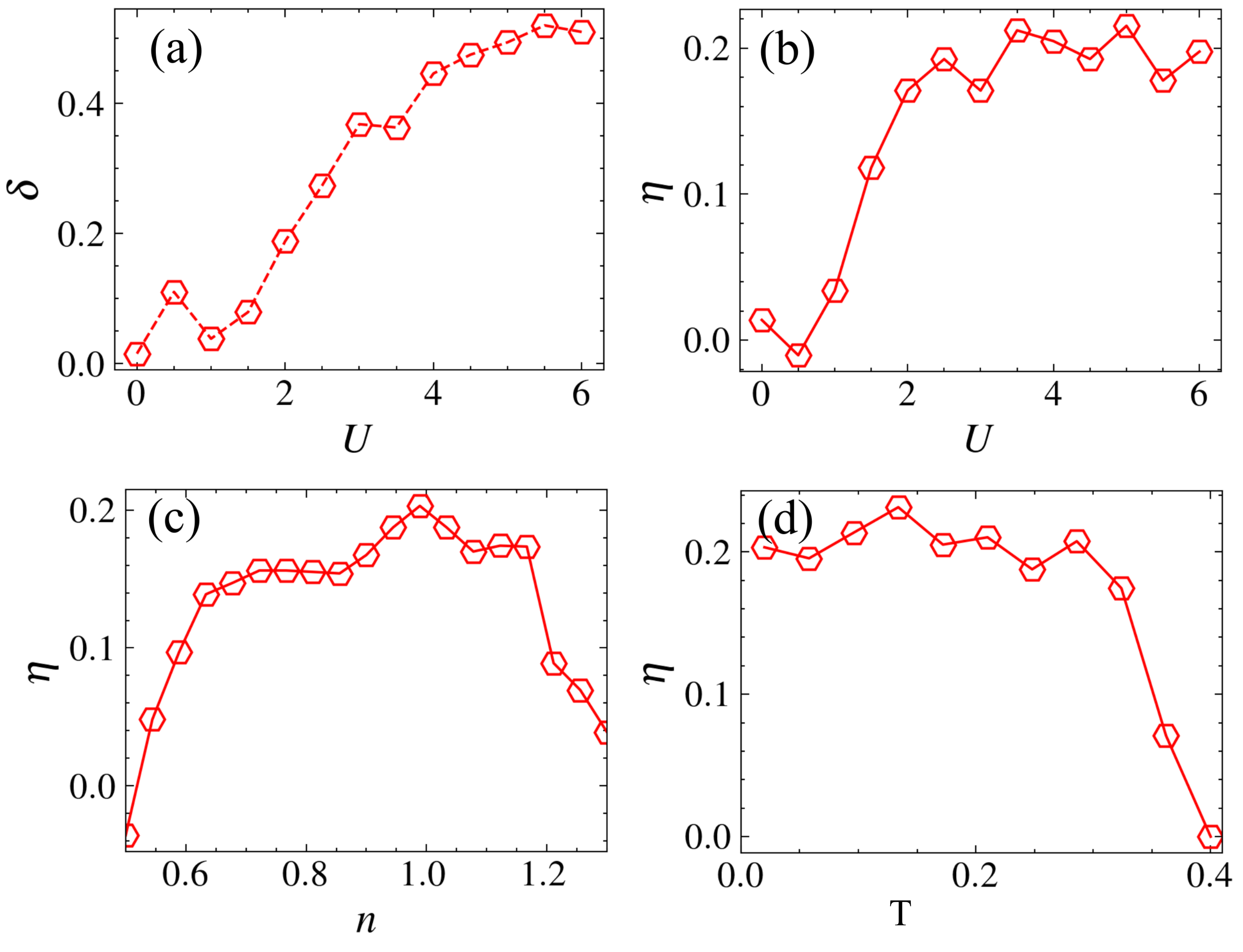}
\caption{(a) The Cooper pair momentum asymmetry quantifier $\delta$ and (b) the superconducting diode efficiency $\eta$ as functions of the HK interaction strength $U$, calculated at a temperature $T=0.03$ and $n=1$. (c) The efficiency $\eta$ as a function of the filling fraction $n$ at $U=4$ and $T=0.03$. (d) The efficiency $\eta$ as a function of temperature $T$ at $U=4$ and half-filling ($n=1$). For all panels, the attractive pairing strength is fixed at $g=1.8$.}
    \label{eta}
\end{figure}

Specifically, we substitute the eigenvalues ${E_{n,k,q}}$ of the $H_{BdG}$ into the partition function in Eq. (\ref{part}) to express the condensation energy as a function of $\Delta_{q}$ and $q$. By minimizing the condensation energy with respect to $\Delta_{q}$ for a fixed momentum $q$, we determine it as a function of $q$ and extract the $J_q$ as a function of $q$.
As shown in Fig. (\ref{freej}), the quality factors are $\eta = 0.046$ for $U = 0$ and $\eta = 0.20$ for $U = 4$, demonstrating that the HK interaction enhances the quality factor. {\bluee{We note the existence of a secondary local minimum in the free energy near $q \approx 1$ [see Fig. \ref{freej}(c)]. While this state possesses its own local critical currents (visible as high-momentum extrema in $J_q$), it has a significantly higher energy than the global ground state near $q \approx 0$. Consequently, in standard equilibrium or DC transport experiments, the system is expected to relax to the global minimum, and the transport properties will be governed solely by the critical currents near $q \approx 0$.

Furthermore, the momenta corresponding to the free energy minima in Fig. \ref{freej}(c) do not strictly coincide with the analytical estimates $q^{(1)}$ and $q^{(2)}$ derived in Sec. II. This quantitative difference arises because the numerical calculation fully accounts for the competition between bands and contributions from states away from the Fermi surface, effects that are treated only approximately in the low-energy effective theory.}}
The dependence of the SDE quality factor $\eta$ on the HK interaction strength $U$, illustrated in Fig. (\ref{eta}), qualitatively agrees with $\delta$. {\bluee{The quality factor $\eta$ exhibits a dip at weak interaction strengths before increasing robustly. At large $U$, the quality factor saturates. This saturation occurs because the HK interaction-induced reconstruction of the Fermi momenta---which underpins the critical momentum asymmetry---stabilizes in the strong coupling limit.
}}

{\bluee{ To further explore the predictive power of our model beyond the proof-of-concept amplification of the SDE, we investigate the dependence of the quality factor on doping and temperature. Figure \ref{eta}(c) illustrates $\eta$ as a function of the filling fraction $n$ at a fixed interaction strength $U=4$ and $T=0.03$. The quality factor exhibits a strong sensitivity to the carrier density, reaching its maximum amplification near half-filling ($n \approx 1$) and gradually diminishing as the system is doped away from this optimal regime. Furthermore, the temperature dependence of $\eta$ at half-filling is shown in Fig. \ref{eta}(d). The quality factor decreases monotonically with increasing temperature until it abruptly drops to zero at the critical temperature $T_c \approx 0.4$. Notably, the vanishing of the SDE coincides precisely with a first-order superconducting phase transition, which is consistent with the anomalous thermodynamic properties induced by the strong momentum-space correlations in the HK model.}}

\section{ Connection to Experiments}

{\bluee{The recent rapid growth in experimental activity observing the SDE in strongly correlated systems---such as magic-angle twisted bilayer graphene \citep{lin2022zero,diez2023symmetry} , FeSe \cite{nagata2025field}, and cuprates \cite{qi2025high}---highlights significant anomalies that traditional weak-coupling theories fail to capture. In standard weakly-interacting paradigms, the SDE quality factor $\eta$ typically scales perturbatively with symmetry-breaking parameters and is generally predicted to be small unless the system is highly fine-tuned. Furthermore, weak-coupling models rely on well-defined Landau quasiparticles. However, in systems like TBG, the SDE is observed to be highly tunable via gating and becomes unusually prominent in regimes of strong electronic correlation, where standard quasiparticle descriptions are heavily modified.

The asymmetric HK model, while an idealized exactly solvable framework, sheds critical light on these anomalies by isolating the role of strong momentum-space correlations. Our results demonstrate that strong electron-electron interactions do not merely compete with or suppress superconductivity, but can actively amplify the nonreciprocal supercurrent. The correlation-induced band splitting redistributes the spectral weight, effectively magnifying the intrinsic band asymmetry at the Fermi level even within the correlated metallic phase. 

Our finding that the SDE quality factor reaches its maximum amplification near half-filling ($n \approx 1$) [as shown in Fig. \ref{eta}(c)] provides a qualitative theoretical bridge to experimental observations. It illustrates one possible mechanism for why nonreciprocal transport signatures can be heavily pronounced at specific carrier densities \citep{lin2022zero,diez2023symmetry}. While realistic materials possess additional complexities such as multi-orbital physics and specific lattice symmetries, the key mechanism driving the enhancement in our model---spectral weight transfer---may be more broadly applicable. }}

\titlespacing*{\section}{0pt}{1cm}{0.2cm}

\section{Conclusion}
\label{CONHK}
In conclusion, we have demonstrated that strong electron correlations, encapsulated by the HK interaction, serve as a powerful mechanism for shaping and amplifying the SDE in systems with intrinsic band asymmetry. Our approach, combining self-consistent numerical simulations with low-energy analysis, reveals that the HK interaction induces a correlation-driven band splitting. This splitting critically enhances the intrinsic asymmetry of the system, leading to a distinct non-monotonic closure of the superconducting energy spectra---the eigenvalues of the BdG Hamiltonian---at inequivalent critical momenta, $q_c^{+} \neq |q_c^{-}|$, for opposing current directions. 

The resulting disparity in critical supercurrents, $J_c^{+} \neq |J_c^{-}|$, marks a significant enhancement of nonreciprocal transport. Our work establishes that this effect is governed by a correlation-tuned asymmetry in the Cooper pair momentum distribution, thereby bridging the gap between weakly-correlated SDE paradigms and the physics of strong coupling. These findings provide a foundational principle for engineering nonreciprocal superconductivity and superconducting diode devices through the targeted manipulation of many-body interactions. 
\acknowledgments
\textit{Acknowledgments}---K.C. acknowledges support from the Fundamental Research Funds for the Central Universities. P.H. acknowledges support from the Department of Energy grant no. DE-SC0022264.

\noindent $^{\dagger}$ KaiChenPhys@tongji.edu.cn
\noindent $^{*}$phosur@central.uh.edu
\bibliography{sample}

\appendix
\section{Full BdG Hmialtonian}
\label{appendix}
In this appendix, we provide a detailed derivation of the BdG Hamiltonian.

To write down the BdG Hamiltonian in the Hilbert space $\mathcal{H}_{k+q}\otimes\mathcal{H}_{-k}$, we examine how $H_{\text{BdG}}$ acts on its basis, as described below
\begin{equation}
\left\{
\begin{aligned}
&H_{\text{BdG}} |0,0\rangle = \sqrt{2} \, |+\rangle, \\
&H_{\text{BdG}} |+\rangle = \left( \xi_{k+q} + \xi_{-k} \right) |+\rangle \\
&\qquad - \sqrt{2} \, \Delta_{q} |\downarrow\uparrow, \downarrow\uparrow\rangle - \sqrt{2} \, \Delta_{q} |0,0\rangle, \\
&H_{\text{BdG}} |\downarrow\uparrow, \downarrow\uparrow\rangle = \left( 2\xi_{k+q} + 2\xi_{-k} + 2U \right) |\downarrow\uparrow, \downarrow\uparrow\rangle \\
&\qquad - \sqrt{2} \, \Delta_{q} |+\rangle.
\end{aligned}
\right.
\label{abasiS3}
\end{equation}
where the state $|+\rangle\equiv \frac{1}{\sqrt{2}}\left(|\uparrow,\downarrow\rangle+ |\downarrow,\uparrow\rangle\right)$ is one of the Bell states with maximum entanglement.
\begin{equation} \begin{aligned}
    H_{\text{BdG}}|\uparrow,\uparrow\rangle&= \left(\xi_{k+q}+\xi_{-k}\right)|\uparrow,\uparrow\rangle,
\end{aligned} \end{equation}
\begin{equation}  \begin{aligned}
H_{\text{BdG}}|\downarrow,\downarrow\rangle &= \left(\xi_{k+q}+\xi_{-k}\right)|\downarrow,\downarrow\rangle,
\end{aligned} \end{equation}
\begin{equation}  \begin{aligned}
H_{\text{BdG}}|0,\uparrow\downarrow\rangle &= \left(2\xi_{-k}+U\right)|0,\uparrow\downarrow\rangle,
\end{aligned} \end{equation}
\begin{equation}  \begin{aligned}
H_{\text{BdG}}|\uparrow\downarrow,0\rangle &= \left(2\xi_{k+q}+U\right)|\uparrow\downarrow,0\rangle,
\end{aligned}  \end{equation}
\begin{equation} \left\{ \begin{aligned}
    H_{\text{BdG}}|0,\uparrow\rangle&= \xi_{-k}|0,\uparrow\rangle-\Delta_{q}|\uparrow,\uparrow\downarrow\rangle, \\
    H_{\text{BdG}}|\uparrow,\uparrow\downarrow\rangle&= \left(\xi_{k+q}+2\xi_{-k}+U\right)|\uparrow,\uparrow\downarrow\rangle-\Delta_{q}|0,\uparrow\rangle
\end{aligned} \right. \end{equation}
\begin{equation} \left\{ \begin{aligned}
    H_{\text{BdG}}|0,\downarrow\rangle&= \xi_{-k}|0,\downarrow\rangle-\Delta_{q}|\downarrow,\uparrow\downarrow\rangle, \\
    H_{\text{BdG}}|\downarrow,\uparrow\downarrow\rangle&= \left(\xi_{k+q}+2\xi_{-k}+U\right)|\downarrow,\uparrow\downarrow\rangle-\Delta_{q}|0,\downarrow\rangle
\end{aligned} \right. \label{abasis2} \end{equation}
\begin{equation} \left\{ \begin{aligned}
    H_{\text{BdG}}|\downarrow,0\rangle&= \xi_{k+q}|\downarrow,0\rangle-\Delta_{q}|\uparrow\downarrow,\downarrow\rangle, \\
    H_{\text{BdG}}|\uparrow\downarrow,\downarrow\rangle&= \left(2\xi_{k+q}+\xi_{-k}+U\right)|\uparrow\downarrow,\downarrow\rangle-\Delta_{q}|\downarrow,0\rangle
\end{aligned} \right. \end{equation}
\begin{equation}
\left\{
\begin{aligned}
    H_{\text{BdG}}|\uparrow,0\rangle&= \xi_{k+q}|\uparrow,0\rangle-\Delta_{q}|\uparrow\downarrow,\uparrow\rangle \\
    H_{\text{BdG}}|\uparrow\downarrow,\uparrow\rangle&= \left(2\xi_{k+q}+\xi_{-k}+U\right)|\uparrow\downarrow,\uparrow\rangle-\Delta_{q}|\uparrow,0\rangle
\end{aligned}
\right.\label{abasiS1}
\end{equation}

As expected, the Hilbert space decouples into sectors with odd and even fermion parity. The even-parity sector comprises one 3-dimensional subspace and four 1-dimensional subspaces, whereas the odd-parity sector consists of four 2-dimensional subspaces. The matrix representation of the BdG Hamiltonian in these subspaces can be directly constructed from Eqs.~(\ref{abasiS3})--(\ref{abasiS1}).

\end{document}